%% file: First_order_GR.tex
\documentclass[11pt,a4paper]{article}
\usepackage{amsmath,amsfonts,amssymb,amscd,graphicx}
\catcode`\@=11
\@addtoreset{equation}{section}
\catcode`\@=12

\newtheorem{Proposition}{Proposition}

\newfont{\gotico}{eufm10 scaled\magstephalf}
\newfont{\qvd}{msam10 scaled\magstephalf}

\def\demo{\par\noindent{\sc Proof. }\begingroup}
\def\enddemo{\hskip1em \mbox{\qvd \char3}\endgroup\par\medskip}

\def\interior{\,\hbox{\vrule depth0pt height.6pt width4pt%
\vrule depth0pt height8pt}\;\,}

\def\de#1/de#2{\frac{\partial {#1}}{\partial {#2}}}
\def\De#1/de#2{\dfrac{\partial {#1}}{\partial {#2}}}

\def\det{{\rm det}\,}

\def\a{\alpha}

\def\jE{{\cal J}\/({\cal E})}
\def\E{{\cal E}}

\begin{document}
\vskip-2cm

\title{A first--order purely frame--formulation of General Relativity}

\author{Stefano Vignolo, Roberto Cianci\\
        DIPTEM Sez. Metodi e Modelli Matematici, Universit\`a di Genova \\
                Piazzale Kennedy, Pad. D - 16129 Genova (Italia)\\ 
                E-mail: vignolo@diptem.unige.it, cianci@diptem.unige.it
\and
       Danilo Bruno \\
        Dipartimento di Matematica, Universit\`a di Genova \\
        Via Dodecaneso 35 - 16146 Genova (Italia) \\
                E-mail: bruno@dima.unige.it
}

\date{}              
\maketitle

\begin{abstract}{In the gauge natural bundle framework a new space is introduced and
a first--order purely frame--formulation of General Relativity is obtained.}
\end{abstract}

\input{par_1}

\input{par_2}

\input{ref.tex}
\end{document}

%% file: par_1.tex
In some of our recent works \cite{CVB1,CVB2,VC} a new geometrical framework
for Yang--Mills field theories and General Relativity in the tetrad--affine formulation
has been developed.

The construction of the new geometrical setting has been obtained quotienting the
first--jet bundles of the configuration spaces of the above theories in a suitable way,
resulting into the introduction of a new family of fiber bundles.

In this letter we show that these new spaces allow a (covariant) first--order purely frame--formulation
of General Relativity.

The whole geometrical construction will be developed within the gauge natural bundle
framework \cite{FF}, which provides the suitable mathematical setting for globally describing
gravity in the tetrad formalism.

To start with, let $M$ be a space--time manifold, allowing a metric tensor $g$
with signature $\eta=(1,3)$: the manifold $M\/$ will be called a $\eta$-manifold and the
metric tensor canonical representation will be $\eta^{\mu\nu}:=diag(-1,1,1,1)$. Moreover,
let $L\/(M)$ be the frame--bundle over $M$ and $P\to M$ a principal fiber bundle over $M$
with structural group $SO\/(1,3)$.

The configuration space of the theory (the tetrad space) is a $GL\/(4,\Re)$
bundle $\pi:\E\to M$, associated to $P\times_M L\/(M)$ through the left--action
\begin{equation}\label{2.1}
    \lambda : (SO\/(1,3)\times GL\/(4,\Re))\times GL\/(4,\Re)\to GL\/(4,\Re),\quad
\lambda\/(\Lambda,J;X)=\Lambda\cdot X\cdot J^{-1}
\end{equation}
Taking eq.~\eqref{2.1} into account, the space $\E$ can be referred to local
fibered coordinates $x^i,e^\mu_i$ $(i,\mu=1\ldots 4)$, undergoing the transformations
laws
\begin{equation}\label{2.2}
    \bar{x}^j = \bar{x}^j\/(x^i),\qquad \bar{e}^\mu_j = e^\sigma_i
\Lambda^\mu_{\;\;\sigma}\/(x) \de x^i/de{\bar{x}^j}
\end{equation}
where $\Lambda^\mu_{\;\;\sigma}\/(x)\in SO\/(1,3)\;\forall\;x\in M$.

Under these circumstances the tetrad fields can be identified with the sections
of the bundle $\E\to M$. It is worth noticing that the conditions making $M$ into a
$\eta$-manifold allow to choose the principal bundle $P$ in such a way that $\E\/$ admits global sections (see
\cite{FF}). In the following, such a choice will be systematically adopted.

Moreover, we also remind that (compare with \cite{FF} again) there exists a
one-to-one correspondence between the global sections of the bundle $\pi:\E\to M$ and the
principal morphisms $i:P\to L\/(M)$. 

Whenever two principal connections
$\omega_{i\;\;\;\nu}^{\;\;\mu}$ over $P$ and $\Gamma_{ih}^{k}$ over $L\/(M)$ are
given, the covariant exterior differential of any tetrad field
$e^\mu\/(x)=e^\mu_i\/(x)\/dx^i$ is well defined as
\begin{equation}\label{2.3}
    D e^\mu := \nabla_j\,e^\mu_{\;i}\/dx^j\wedge dx^i
\end{equation}
where
\[
\nabla_j\,e^\mu_{\;i} = \de e^\mu_{\;i} /de{x^j} + \omega_{j\;\;\nu}^{\;\;\mu}
e^\nu_{\;i} - \Gamma_{ji}^{k} e^\mu_{\;k}
\]
The first jet bundle associated to the fibration $\pi:\E\to M$ is now taken
into account. A set of jet--coordinates over $j_1\/(\E)$ is provided by
$x^i,e^\mu_i,e^\mu_{\;ij}\left(\simeq \de e^\mu_i /de{x^j}\right)$, subject to the
transformation laws \eqref{2.2} together with
\begin{equation}\label{2.4}
\bar{e}^\mu_{jk} = e^\sigma_{ih}\de{x^h}/de{\bar{x}^k} \Lambda^\mu_{\;\;\sigma}\de
x^i/de{\bar{x}^j} + e^\sigma_i \de\Lambda^\mu_{\;\;\sigma}/de{x^h}\de
x^h/de{\bar{x}^k}\de x^i/de{\bar{x}^j} +
e^\sigma_i\Lambda^\mu_{\;\;\sigma}\frac{\partial^2 x^i}{\bar{x}^k \bar{x}^j}
\end{equation}
The frame--formulation of general relativity that we propose here is based on
the introduction of the following equivalence relation on $j_1\/(\E)$. Let
$z=(x^i,e^\mu_i,e^\mu_{\;ij})$ and $\hat z =(x^i,\hat e^\mu_i,\hat e^\mu_{\;ij})$ be two
elements of $j_1\/(\E)$, chosen in such a way that they have the same projection over
$M$, namely $\hat \pi\/(z)=\hat \pi\/(\hat z)=x$, with $\hat \pi:j_1\/(\E)\to M$. We
denote by $e^\mu\/$ and $\hat e^\mu\/$ two different sections of the bundle
$\pi:\E\to M$, respectively chosen among the representatives of the equivalence classes
$z$ and $\hat z$. Then, we make $z$ equivalent to $\hat z$ if and only if
\begin{equation}\label{2.5}
    e^\mu\/(x) = \hat e^\mu\/(x) \quad {\rm and } \quad D e^\mu\/(x)= D \hat e^\mu\/(x)
\end{equation}
for every choice of a principal connection $\omega$ on $P$ and $\Gamma$ on $L\/(M)$.

It is easy to see that $z\sim\hat z$ if and only if the following local
coordinates expression holds:
\begin{equation}\label{2.6}
    e^\mu_i =  \hat e^\mu_i \quad {\rm and } \quad (e^\mu_{ij} - e^\mu_{ji}) =
    (\hat e^\mu_{ij} - \hat e^\mu_{ji})
\end{equation}
We denote by $\jE$ the quotient space $\jE:=j_1\/(E)/\sim$ and by
$\rho:j_1\/(E)\to\jE$ the corresponding quotient canonical projection. A system of local
fibered coordinates on the bundle $\jE$ is provided by
$x^i,e^\mu_i,E^\mu_{\;ij}:=\frac{1}{2}\left(e^\mu_{ij} - e^\mu_{ji} \right) (i<j)$,
subject to the transformation laws \eqref{2.2}, together with:
\begin{equation}\label{2.7}
\bar{E}^\mu_{jk} = E^\sigma_{ih}\Lambda^\mu_{\;\;\sigma}\de x^h/de{\bar{x}^k}\de
x^i/de{\bar{x}^j} + \frac{1}{2}e^\sigma_i\de\Lambda^\mu_{\;\;\sigma}/de{x^h}\de
x^h/de{\bar{x}^k}\de x^i/de{\bar{x}^j} -
\frac{1}{2}e^\sigma_i\de\Lambda^\mu_{\;\;\sigma}/de{x^h}\de x^h/de{\bar{x}^j}\de
x^i/de{\bar{x}^k}
\end{equation}
The geometry of the quotient space $\jE$ has been deeply examined in some
previous papers \cite{CVB1,CVB2,VC}. As a matter of fact, the quotient projection endows
the bundle $\jE$ of  most of the standard features of jet--bundles geometry. The
principal results are shortly reported below (see \cite{CVB1,CVB2,VC} for a more detailed
discussion).

\bigskip
\noindent $\bullet$ {\it ${\cal J}$-extension of sections.} The ${\cal J}$-extension of a
section $\sigma:M\to \E$ is defined as ${\cal J}\sigma := \rho\circ j_1\/\sigma$, namely
projecting the jet--extension $j_1\/(\sigma)$ on $\jE$ by means of the quotient
projection $\rho$. A section $\gamma:M\to\jE$ is said {\it holonomic} if there exists a
section $\sigma:M\to \E$ such that $\gamma = {\cal J}\sigma$. In local coordinates, a
section $\gamma$ is holonomic if and only if $\gamma : x \mapsto \left(
x^i,e^\mu_i\/(x),E^\mu_{ij}\/(x) = \frac{1}{2} \left( \de e^\mu_{i}\/(x) /de{x^j} - \de
e^\mu_{j}\/(x) /de{x^i}\right)\right)$.

\bigskip
\noindent $\bullet$ {\it Contact forms}. Let us define the following $2$-form on $\jE$:
\begin{equation}\label{2.7c}
\theta^\mu := de^\mu_i \wedge dx^i + E^\mu_{ij} dx^i\wedge dx^j
\end{equation}
where $E^\mu_{ij}:=-E^\mu_{ji}\/$ whenever $i>j\/$. Under a change of local coordinates \eqref{2.2} and \eqref{2.7}, the $2$-forms
\eqref{2.7c} undergo the transformation laws
\begin{equation}\label{2.7cc}
    \bar \theta^\mu = \Lambda^\mu_{\;\;\nu} \theta^\nu
\end{equation}
The vector bundle which is locally spanned by the $2$-forms \eqref{2.7c} will
be called the contact bundle ${\cal C}(\jE)$ and any section $\eta:\jE\to {\cal C}(\jE)$
will be called a contact $2$-form. Contact forms $\eta\/$ are such that $\gamma^*\/(\eta)=0$
whenever $\gamma:M\to \jE$ is holonomic. Conversely, if a section $\gamma: M\to\jE\/$ is such that
$\gamma^*\/(\eta)=0\/$ for all contact forms $\eta\/$, then $\gamma\/$ is holonomic.

\bigskip
\noindent $\bullet$ {\it $\cal{J}$-prolongations of morphisms and vector fields}. A
suitable family of morphisms $\Phi:\E\to \E$, fibered over $M$, can be raised to a family of morphisms ${\cal J}
\Phi : \jE \to \jE$ considering their ordinary jet--prolongations and projecting them to
$\jE$ through the quotient map, namely:
\[
{\cal J} \Phi (z) := \rho \circ j_1\Phi\/(w) \quad \forall\; w \in \rho^{-1}\/(z)\;,\;
z\in \jE
\]
In order that the above definition makes sense, such morphisms $\Phi:\E\to \E$ have to satisfy the condition:
\begin{equation}\label{2.7x}
\rho\circ j_{1}\Phi\/(w_1)=\rho\circ j_{1}\Phi\/(w_2) \qquad\forall\;w_1,w_2 \in
\rho^{-1}\/(z)\;,\;
z\in \jE
\end{equation}
Referring to \cite{CVB1} for the proof, it is easy to see that the only morphisms
satisfying condition \eqref{2.7x} are necessarily of the form:
\begin{equation}\label{2.7xx}
\left\{
\begin{array}{l}
y^i = \chi^i\/(x^j)\\
\\
\hat e^\nu_i = \Phi^\nu_i\/(x^j,e^\mu_j)= \Gamma^\nu_\mu\/(x) \De x^r/de{y^i} e^\mu_r +
f^\nu_i\/(x)
\end{array}
\right.
\end{equation}
where $\Gamma^\nu_\mu\/(x)\/ $ and $f^\nu_i\/(x)\/$ are arbitrary local functions on
$M\/$. Their ${\cal J}-$prolongation is:
\[
\left\{
\begin{array}{l}
y^i = \chi^i\/(x^k)\\
\\
\hat e^\nu_i = \Gamma^\nu_\mu\/(x) \De x^r/de{y^i} e^\mu_r + f^\nu_i\/(x)\\
\\
\hat E^\nu_{ij} = \Gamma^\nu_{\mu}E^\mu_{ks}\De x^k/de{y^i}\De x^s/de{y^j} +
\frac{1}{2}\left[\De\Gamma^\nu_\mu/de{x^k}\left(\De x^k/de{y^j}\De x^r/de{y^i} - \De
x^k/de{y^i}\De x^r/de{y^j}\right)e^\mu_r + \De f^\nu_i/de{x^k}\De x^k/de{y^j} - \De
f^\nu_j/de{x^k}\De x^k/de{y^i}\right]\end{array} \right.
\]
In a similar way (compare with \cite{CVB1}), it is easy to prove that the only vector
fields of the form
\begin{equation}\label{2.7xxx}
X=\epsilon^i\/(x^j)\,\de /de{x^i} +
\left(-\de{\epsilon^k}/de{x^q}e^\mu_k + D^\mu_\nu\/(x^j)e^\nu_q +
G^\mu_q\/(x^j)\right)\,\de /de{e^\mu_q}
\end{equation}
where $\epsilon^i$, $D^\mu_\nu$ and $G^\mu_q$ are arbitrary local functions on $M$, can be
${\cal J}-$prolonged to vector fields over $\jE$ as follows:
\begin{equation}\label{2.7xxxx}
{\cal J}\/(X)\/(z):=\rho_{* \rho^{-1}\/(z)}(j_1\/(X)) \qquad\forall\, z\in {\cal J}\/(\E)
\end{equation}
The resulting vector field has the form:
\[
{\cal J}\/(X)=\epsilon^i\/(x^j)\,\de /de{x^i} + \left(-\de{\epsilon^k}/de{x^q}e^\mu_k +
D^\mu_\nu\/(x^j)e^\nu_q + G^\mu_q\/(x^j)\right)\,\de /de{e^\mu_q} +
\sum_{i<j}h^\mu_{ij}\,\de /de{E^\mu_{ij}}
\]
where
\[
h^\mu_{ij}= \frac{1}{2}\left( \de D^\mu_\nu/de{x^j}e^\nu_i - \de D^\mu_\nu/de{x^i}e^\nu_j
+ \de G^\mu_i/de{x^j} - \de G^\mu_j/de{x^i}\right) + D^\mu_{\nu}E^\nu_{ij} + \left(
E^\mu_{ki}\de{\epsilon^k}/de{x^j} - E^\mu_{kj}\de{\epsilon^k}/de{x^i}\right)
\]
In the following discussion the central role will be played by a specific
coordinate transformation in the space $\jE$. More precisely, the main idea consists
in choosing the components of the {\it spin--connections} generated by the tetrads
themselves as fiber coordinates on the bundle $\jE$. 

To see this point, let $z=(x^i,e^\mu_i,E^\mu_{\;ij})$
be an element of $\jE$, $x=\hat\pi\/(z)$ its projection over $M$ and $e^\mu$ a
representative tetrad belonging to the equivalence class $z$. Moreover, if
$g=\eta_{\mu\nu} e^\mu\otimes e^\nu$ is the metric on $M$ induced by the tetrad $e^\mu$,
denote by $\Gamma_{ih}^k$ its associated Levi--Civita connection. The latter is a
principal connection on $L\/(M)$ and can be pulled--back to a spin--connection
$\omega_{i\;\;\;\nu}^{\;\;\mu}$ over $P$ by means of the tetrad $e^\mu$ itself (i.e.
through the principal morphism $i:P\to L\/(M)$ associated to the tetrad $e^\mu$).

The relation between the coefficients $\Gamma_{ih}^k$ of the Levi--Civita
connection and the coefficients $\omega_{i\;\;\;\nu}^{\;\;\mu}$ of the associated spin--connection,
evaluated at the point $x=\hat \pi\/(z)\in M$, is expressed by the equation
\begin{equation}\label{2.8}
\omega^{\;\;\mu}_{i\;\;\;\nu}\/(x) = e^\mu_k\/(x)\left( \Gamma^k_{ij}e^j_\nu\/(x) +
\de{e^k_\nu\/(x)}/de{x^i} \right)
\end{equation}

In other and simpler words, the latter can be though as the Levi--Civita connection
expressed in terms of the non--holonomic basis $e^\mu\/(x)$. If the coefficients
$\Gamma_{ih}^k$ are written in terms of the tetrad $e^\mu$ and its derivatives, one gets
the well--known expression
\begin{equation}\label{2.9}
\omega^{\;\;\mu}_{i\;\;\nu}\/(x):= e^\mu_p\/(x) \left( \Sigma^p_{\;\;ji}\/(x) -
\Sigma_{j\;\;i}^{\;\;p}\/(x) + \Sigma_{ij}^{\;\;\;p}\/(x) \right) e^j_{\nu}\/(x)
\end{equation}
where
\begin{equation}\label{2.10}
\Sigma^p_{\;\;ji}\/(x):= e^p_\lambda\/(x) E^\lambda_{ij}\/(x) =
e^p_\lambda\/(x)\frac{1}{2}\left( \de{e^\lambda_i\/(x)}/de{x^j} -
\de{e^\lambda_j\/(x)}/de{x^i} \right)
\end{equation}
the Latin indexes being lowered and raised by means of the metric $g=\eta_{\mu\nu}e^\mu \otimes e^\nu\/$.

Equations \eqref{2.9} and \eqref{2.10} show that the values of the coefficients
of the spin--connection $\omega_{i\;\;\;\nu}^{\;\;\mu}$, evaluated in $x=\hat \pi\/(z)$,
are independent of the choice of the representative $e^\mu$ in the equivalence class
$z\in \jE$.

Moreover, the torsion--free condition for the connection
$\omega_{i\;\;\;\nu}^{\;\;\mu}$ gives a sort of inverse relation of eq.~\eqref{2.9} in the form
\begin{equation}\label{2.11}
2E^\mu_{ij}\/(x) = \de{e^\mu_i\/(x)}/de{x^j} -
\de{e^\mu_j\/(x)}/de{x^i} = \omega^{\;\;\mu}_{i\;\;\;\nu}\/(x)e^\nu_j\/(x) -
\omega^{\;\;\mu}_{j\;\;\;\nu}\/(x)e^\nu_i\/(x)
\end{equation}

Because of the metric compatibility condition $\omega_i^{\;\;\mu\nu}:=
\omega_{i\;\;\;\sigma}^{\;\;\mu} \eta^{\sigma\nu} = - \omega_i^{\;\;\nu\mu}$, there exists
a one-to-one correspondence between the values of the antisymmetric part of the
derivatives $E^\mu_{ij}\/(x) = \frac{1}{2} \left( \de e^\mu_{i}\/(x) /de{x^j} - \de e^\mu_{j}\/(x)
/de{x^i}\right)$ and the coefficients of the spin--connection $\omega_i^{\;\;\mu\nu}\/(x)$ in
the point $x=\hat\pi\/(z)$.

The above considerations allow us to take the quantities
$\omega_i^{\;\;\mu\nu}$ as fiber coordinates of the bundle $\jE$, looking at the
relations \eqref{2.9} and \eqref{2.11} as coordinate changes in $\jE$.

It is a well--known fact that the coordinate transformations \eqref{2.2} induce
the following transformation laws for the spin--connection coefficients
$\omega_i^{\;\;\mu\nu}$:
\begin{equation}\label{2.12}
\bar{\omega}^{\;\;\mu\nu}_{i}=\Lambda^\mu_{\;\;\sigma}\/(x)\Lambda^{\nu}_{\;\;\gamma}\/(x)\de
x^j/de{\bar{x}^i}\omega^{\;\;\sigma\gamma}_{j} - \Lambda_{\sigma}^{\;\;\eta}\/(x)\de
\Lambda^\mu_{\;\;\eta}\/(x)/de{x^h}\de x^h/de{\bar{x}^i}\eta^{\sigma\nu}
\end{equation}
where $\Lambda_{\sigma}^{\;\;\nu}:=\Lambda^{\alpha}_{\;\;\beta}\eta_{\alpha\sigma}\eta^{\beta\nu}=\left(\Lambda^{-1}\right)^\nu_{\;\;\sigma}\/$.

%% file: par_2.tex
Let us now define the variational principle from which we shall deduce the field equations 
for General Relativity directly on the only manifold ${\cal J}\/(\cal E)\/$.

To this end, we first introduce the $4$-form on ${\cal J}\/(\cal E)\/$ locally
described as
\begin{equation}\label{3.1}
\Theta := \frac{1}{4}\epsilon^{qpij}\epsilon_{\mu\nu\lambda\sigma}\/e^\mu_{q}e^\nu_{p}\left( d\omega_{i}^{\;\;\lambda\sigma} \wedge ds_j + \omega_{j\;\;\;\eta}^{\;\;\lambda}\omega_{i}^{\;\;\eta\sigma}\,ds \right)
\end{equation}
where $ds:=dx^1 \wedge\ldots\wedge dx^4\/$, $ds_i := \de /de{x^i}\interior ds\/$ and $\epsilon\/$ denotes the Levi--Civita permutation symbol. The following result holds true
\begin{Proposition}\label{Pro3.1}
The form $\Theta\/$ (\ref{3.1}) is invariant under the coordinate transformations
\eqref{2.2}, \eqref{2.12} on the manifold ${\cal J}\/(\cal E)\/$.
\end{Proposition}
\demo It is a direct check, taking eqs.~\eqref{2.2}, \eqref{2.12} and the identities
\[
\begin{split}
\bar{\omega}_{j\;\;\;\eta}^{\;\;\tau}\bar{\omega}_{i\;\;\;\sigma}^{\;\;\eta} = \Lambda^{\tau}_{\;\;\alpha}\Lambda_{\sigma}^{\;\;\beta}\de{x^k}/de{\bar{x}^j}\de{x^h}/de{\bar{x}^i}\omega_{k\;\;\;\lambda}^{\;\;\alpha}\omega_{h\;\;\;\beta}^{\;\;\lambda} +
\Lambda^{\tau}_{\;\;\alpha}\de{\Lambda_{\sigma}^{\;\;\beta}}/de{x^h}\de{x^h}/de{\bar{x}^i}\de{x^k}/de{\bar{x}^j}\omega_{k\;\;\;\beta}^{\;\;\alpha} +\\
-\de{\Lambda_{\;\;\alpha}^{\tau}}/de{x^h}\de{x^h}/de{\bar{x}^j}\Lambda_{\sigma}^{\;\;\beta}\de{x^k}/de{\bar{x}^i}\omega_{k\;\;\;\beta}^{\;\;\alpha} -
\de{\Lambda_{\;\;\eta}^{\tau}}/de{x^h}\de{x^h}/de{\bar{x}^j}\de{\Lambda_{\sigma}^{\;\;\eta}}/de{x^k}\de{x^k}/de{\bar{x}^i}
\end{split}
\]
explicitly into account.
\enddemo
Being the form $\Theta\/$ a covariant geometrical object, it can be used to define a
variational problem on the bundle ${\cal J}\/(\cal E)\/$, consisting in the study of the
stationarity conditions for the functional
\begin{equation}\label{3.2}
A\/(\gamma):= \int_D \gamma^*\/(\Theta)
\end{equation}
for every section $\gamma : D\subset M\to {\cal J}\/(\cal E)\/$, $D\/$ compact domain.

The procedure is well known: we take a vertical vector field $X\/$ (with
respect to the fibration ${\cal J}\/({\cal E})\to M\/$) into account and denote by
$\Phi_\xi\/$ its flow; then, we deform any given section $\gamma:M\to \jE\/$ along $X\/$
by setting $\gamma_\xi := \Phi_\xi \circ \gamma\/$. We name first variation of $A\/$ at
$\gamma\/$ in the direction $X\/$ the expression (see, for example, \cite{Hermann})
\begin{equation}\label{3.3}
\frac{\delta A}{\delta X}\/(\gamma) := \frac{d}{d\xi}{\int_D \gamma_\xi^*\/(\Theta)}_{\big|_{\xi =0}}= \int_D \gamma^*\/(X\interior d{\Theta}) + \int_{\partial D} \gamma^*\/(X\interior {\Theta})
\end{equation}
Finally, we look for sections $\gamma:x\to (x^i,e^\mu_i\/(x),\omega_i^{\;\;\mu\nu}\/(x))\/$ ({\it critical points\/}) obeying the ansatz $\frac{\delta A}{\delta X}\/(\gamma)=0\/$, for all compact domains $D\/$ and all infinitesimal deformations $X\/$ vanishing on the boundary $\partial D\/$.

According to eq.~(\ref{3.3}) and to the imposed boundary condition, a section $\gamma\/$ is critical if and only if it
satisfies the equation
\begin{equation}\label{3.4}
\gamma^*\/(X\interior\Theta)=0
\end{equation}
for every vector field $X= X^\mu_i\,\de /de{e^\mu_i} + \frac{1}{2}X_i^{\mu\nu}\,\de
/de{\omega_{i}^{\;\;\mu\nu}}\/$ on ${\cal J}\/(\cal E)\/$ (with $X_i^{\mu\nu}=-X_i^{\nu\mu}\/$ when $\mu > \nu\/$).

In order to make eq.~(\ref{3.4}) explicit, we calculate the differential of the
form $\Theta\/$, namely
\begin{equation}\label{3.5}
d\Theta = \frac{1}{2}\epsilon^{qpij}\epsilon_{\mu\nu\lambda\sigma}e^\mu_q\,de^\nu_{p}\wedge\left( d\omega_{i}^{\;\;\lambda\sigma} \wedge ds_j + \omega_{j\;\;\;\eta}^{\;\;\lambda}\omega_{i}^{\;\;\eta\sigma}\,ds \right) + \frac{1}{2}\epsilon^{qpij}\epsilon_{\mu\nu\lambda\sigma}\/e^\mu_{q}e^\nu_{p}\omega_{j\;\;\;\eta}^{\;\;\lambda}\,d\omega_{i}^{\;\;\eta\sigma} \wedge ds
\end{equation}
\begin{Proposition}\label{Pro3.2}
The following identities
\begin{equation}\label{3.4bis}
\epsilon^{qpij}\epsilon_{\mu\nu\lambda\sigma}\/e^\mu_{q}e^\nu_{p}\omega_{j\;\;\;\eta}^{\;\;\lambda}\,d\omega_{i}^{\;\;\eta\sigma} = - \epsilon^{qpij}\epsilon_{\mu\rho\lambda\sigma}\/e^\mu_{q}e^\nu_{p}\omega_{j\;\;\;\nu}^{\;\;\rho}\,d\omega_{i}^{\;\;\lambda\sigma}
\end{equation}
hold true.
\end{Proposition}
\demo Observing that the expressions $\epsilon^{qpij}\/e^\mu_{q}e^\nu_{p}\/$ are
antisymmetric in the indexes $\mu\/$ and $\nu\/$, the identities \eqref{3.4bis} will be
proved if we can show that the antisymmetric combinations (still in the indexes $\mu\/$
and $\nu\/$) of the $1$-forms
$\epsilon_{\mu\nu\lambda\sigma}\/\omega_{j\;\;\;\eta}^{\;\;\lambda}\,d\omega_{i}^{\;\;\eta\sigma}\/$
and
$-\epsilon_{\mu\rho\lambda\sigma}\/\omega_{j\;\;\;\nu}^{\;\;\rho}\,d\omega_{i}^{\;\;\lambda\sigma}\/$
coincide. In turn, the last assertion is mathematically equivalent to the fact that the
following identities
\begin{equation}\label{3.4tris}
\epsilon^{\mu\nu\alpha\beta}\epsilon_{\mu\nu\lambda\sigma}\/\omega_{j\;\;\;\eta}^{\;\;\lambda}\,d\omega_{i}^{\;\;\eta\sigma}=
-\epsilon^{\mu\nu\alpha\beta}\epsilon_{\mu\rho\lambda\sigma}\/\omega_{j\;\;\;\nu}^{\;\;\rho}\,d\omega_{i}^{\;\;\lambda\sigma}
\end{equation}
hold true. Due to the traceless property $\omega_{i\;\;\;\mu}^{\;\;\mu}=0\/$, a direct
calculation shows that both left and right hand sides of \eqref{3.4tris} are actually equal to
$2\/\left( \omega_{j\;\;\;\eta}^{\;\;\alpha}\,d\omega_i^{\;\;\eta\beta} -
\omega_{j\;\;\;\eta}^{\;\;\beta}\,d\omega_i^{\;\;\eta\alpha} \right)\/$.
\enddemo

Making use of the identities (\ref{3.4bis}), we can rewrite the expression
(\ref{3.5}) in the form
\begin{equation}\label{3.6}
d\Theta = \frac{1}{2}\epsilon^{qpij}\epsilon_{\mu\nu\lambda\sigma}e^\mu_{q}\,de^\nu_{p}\wedge\left( d\omega_{i}^{\;\;\lambda\sigma} \wedge ds_j + \omega_{j\;\;\;\eta}^{\;\;\lambda}\omega_{i}^{\;\;\eta\sigma}\,ds \right) -
\frac{1}{2}\epsilon^{qpij}\epsilon_{\mu\rho\lambda\sigma}\/e^\mu_{q}e^\nu_{p}\omega_{j\;\;\;\nu}^{\;\;\rho}\,d\omega_{i}^{\lambda\sigma} \wedge ds
\end{equation}
Now, given a vector field $X= X^\mu_i\,\de /de{e^\mu_i} + \frac{1}{2}X^{\mu\nu}_i\,\de
/de{\omega_{i}^{\;\;\mu\nu}}\/$ on ${\cal J}\/(\cal E)\/$, we easily have
\begin{equation}\label{3.7}
\begin{split}
X\interior d\Theta = \frac{1}{2}\epsilon^{qpij}\epsilon_{\mu\nu\lambda\sigma}\/e^\mu_{q}\,\left( d\omega_{i}^{\;\;\lambda\sigma} \wedge ds_j + \omega_{j\;\;\;\eta}^{\;\;\lambda}\omega_{i}^{\;\;\eta\sigma}\,ds \right)\/X^\nu_{p} +\\
- \frac{1}{2}\epsilon^{qpij}\epsilon_{\mu\nu\lambda\sigma}\/e^\mu_{q}\,\left( de^\nu_p \wedge ds_j + e^\rho_{p}\omega_{j\;\;\;\rho}^{\;\;\nu}\,ds \right)\/X^{\lambda\sigma}_{i}
\end{split}
\end{equation}
In conclusion, the imposition of condition (\ref{3.4}) yields two sets of final equations
\begin{subequations}
\begin{equation}\label{3.8a}
\epsilon^{qpij}\epsilon_{\mu\nu\lambda\sigma}\/e^\mu_q\/\left( \de{e^\nu_p}/de{x^j} + \omega_{j\;\;\;\rho}^{\;\;\nu}e^\rho_{p} \right) = 0
\end{equation}
\begin{equation}\label{3.8b}
\frac{1}{2}\epsilon^{qpij}\epsilon_{\mu\nu\lambda\sigma}\/e^\mu_q\/\left( \de{\omega_{i}^{\;\;\lambda\sigma}}/de{x^j} + \omega_{j\;\;\;\eta}^{\;\;\lambda}\omega_{i}^{\;\;\eta\sigma} \right) = 0
\end{equation}
\end{subequations}
clearly equivalent to Einstein equations (provided that $\det (e^\mu_{i}) \not = 0\/$).
Indeed, eqs.~(\ref{3.8a}) ensure the kinematic admissibility (the holonomy) of the
critical section $\gamma\/$, namely
\[
2E^\nu_{pj}\/(x) = \omega_{p\;\;\;\rho}^{\;\;\nu}\/(x)e^\rho_{j}\/(x) - \omega_{j\;\;\;\rho}^{\;\;\nu}\/(x)e^\rho_{p}\/(x) = \de{e^\nu_p}/de{x^j}\/(x) - \de{e^\nu_j}/de{x^p}\/(x)
\]
so that the quantities $\omega_{i\;\;\;\nu}^{\;\;\mu}\/(x)\/$ identify with the
coefficients of the spin connection associated to the Levi--Civita connection induced by
the metric $g_{ij}\/(x)=\eta_{\mu\nu}e^\mu_{i}\/(x)e^\nu_{j}\/(x)\/$. Therefore
eqs.~(\ref{3.8b}) are identical to
\[
\frac{1}{4}\epsilon^{qpij}\epsilon_{\mu\nu\lambda\sigma}\/e^\mu_q\/(x)R_{ji}^{\;\;\;\lambda\sigma}\/(x)=0
\]
$R_{ji}^{\;\;\;\lambda\sigma}\/(x):= \de{\omega_{i}^{\;\;\lambda\sigma}}/de{x^j}\/(x) - \de{\omega_{j}^{\;\;\lambda\sigma}}/de{x^i}\/(x) +
\omega_{j\;\;\;\eta}^{\;\;\lambda}\/(x)\omega_{i}^{\;\;\eta\sigma}\/(x) - \omega_{i\;\;\;\eta}^{\;\;\lambda}\/(x)\omega_{j}^{\;\;\eta\sigma}\/(x)\/$ denoting the curvature tensor of the metric $g\/$.

It is worth noticing that the restriction regarding the verticality of the
infinitesimal deformations $X\/$ can be removed, since condition \eqref{3.4}
automatically implies $\gamma^*\/(X\interior\Theta)=0\/$ $\forall\;X\in D^1\/({\cal
J}\/({\cal E}))\/$.

This last fact is important in order to extend the study of Noether vector fields,
conserved currents  and symmetries to the present geometrical setting. In particular, any
given vector field $Z\/$ on ${\cal J}\/({\cal E})\/$ will be called a Noether vector
field if it satisfies the ansatz
\begin{equation}
L_{Z}\Theta = \omega + d\alpha
\end{equation}
where $\omega\/$ is a $4$-form belonging to the ideal generated by the contact forms and
$\a\/$ is any $3$-form on ${\cal J}\/({\cal E})\/$. If $Z\/$ satisfies the trivial case
$L_{Z}{\Theta}_L =0\/$ and projects to $M\/$, then $Z\/$ is an infinitesimal dynamical
symmetry (namely, its flow drags critical sections into as many critical sections). It is
also easy to verify that whenever a Noether vector field $Z\/$ is a ${\cal
J}$-prolongation of some vector field (\ref{2.7xxx}) on ${\cal E}\/$, it again results
into an infinitesimal dynamical symmetry.

Moreover, a corresponding conserved current is always associated with any Noether vector
field $Z\/$ . In fact, given a critical section $\gamma\/$, one has
\begin{equation}
d\gamma^*\/\left( Z\interior{\Theta} - \a\right) = \gamma^*\/\left( \omega - Z\interior d{\Theta}\right) =0
\end{equation}
The current $\gamma^*\/\left( Z\interior{\Theta} - \a\right)\/$ is then conserved on shell.

We conclude this letter by noticing that a new geometrical description of the combined
theory of gravitation and Yang--Mills fields within the framework of $\cal J$-bundles can
be obtained, joining the present geometrical approach with the one developed in
\cite{CVB1,CVB2}. The matter is straightforward and follows the lines already illustrated
in \cite{VC} for the tetrad--affine formulation; for brevity reasons, we leave the
details to the reader.